\documentclass[twocolumn]{jpsj2} 
\usepackage{color}

\def\runauthor{C. Yasuda, S. Todo and H. Takayama}

\title{Bond-Dilution-Induced Quantum Phase Transitions in 
Heisenberg Antiferromagnets}

\author{Chitoshi \textsc{Yasuda}$^{1}$\thanks{E-mail address:
cyasuda@phys.aoyama.ac.jp}, Synge \textsc{Todo}$^{2,3}$,
and Hajime \textsc{Takayama}$^{4}$}

\inst{$^{1}$Department of Physics and Mathematics, Aoyama Gakuin University,
Sagamihara 229-8558, Japan \\
$^{2}$Department of Applied Physics, University of Tokyo, Tokyo
113-8656, Japan \\
$^{3}$CREST, Japan Science and Technology Agency, Kawaguchi, 332-0012,
Japan \\
$^{4}$Institute for Solid State Physics, University of Tokyo, Kashiwa
277-8581, Japan}

\abst{Bond-dilution effects on the ground state of the square-lattice
antiferromagnetic Heisenberg model, consisting of coupled
bond-alternating chains, are investigated by means of the quantum Monte
Carlo simulation.  It is found that, when the ground state of the
non-diluted system is a non-magnetic state with a finite spin gap, a
sufficiently weak bond dilution induces a disordered state with a mid
gap in the original spin gap, and under a further stronger bond dilution
an antiferromagnetic long-range order emerges.  While the
site-dilution-induced long-range order is induced by an infinitesimal
concentration of dilution, there exists a finite critical concentration
in the case of bond dilution.  We argue that this essential difference
is due to the occurrence of two types of effective interactions between
induced magnetic moments in the case of bond dilution, and that the
antiferromagnetic long-range-ordered phase does not appear until the
magnitudes of the two interactions become comparable.}

\recdate{\today}

\kword{dilution-induced antiferromagnetic long-range order, quantum
Monte Carlo simulation, quantum phase transition, non-magnetic impurity,
spin-Peierls compound}

\begin{document}
\maketitle

\section{Introduction} 

Randomness effects on quantum antiferromagnetic (AF) Heisenberg models
with a spin-gapped ground state have attracted much interest in relation
to the impurity-induced AF long-range order (LRO) observed
experimentally, e.g., in the first inorganic spin-Peierls compound
CuGeO$_3$~\cite{hase,regnault}, the two-leg ladder compound
SrCu$_2$O$_3$~\cite{azuma}, and the Haldane compound
PbNi$_2$V$_2$O$_8$~\cite{uchiyama}. In the spin-Peierls compound
CuGeO$_3$, the lattice dimerized state associated with a formation of
spin gap is realized at low temperatures~\cite{hase}. When a small
amount of non-magnetic impurities Zn or Mg are substituted for Cu, the
spin-gapped ground state of the pure system changes to an AF
LRO state~\cite{martin,masuda,manabe}, thereby the
lattice dimerization is preserved. Such an impurity-induced AF LRO has
been observed also in the bond-disorder system,
CuGe$_{1-x}$Si$_x$O$_3$~\cite{regnault,masuda2,kikuchi}.  In early
theoretical and numerical studies on quasi-one-dimensional
diluted Heisenberg antiferromagnets, the effect of inter-chain
interaction is often taken into account by making use of mean-field
approximations, where the site and bond dilutions are
unable to be distinguished with each other.~\cite{miyashita,eggert,fukuyama} By the
recent numerical analyses, however, the characteristic feature of the
site-dilution-induced AF LRO has been understood more clearly by
treating the inter- and intra-chain interactions on an equal
footing.~\cite{imada,wessel,yasuda} More recently we have carried out
extensive numerical simulations also on the bond-diluted system, and
have found that the mechanism for the AF LRO to appear in this system is
essentially different from that in the site-diluted system.  The purpose
of the present paper is to discuss our results in more detail, which
have been reported in refs.~\citen{yasuda2} and \citen{yasuda3} briefly.

\begin{figure}[t]
  \centerline{\resizebox{0.45\textwidth}{!}{\includegraphics{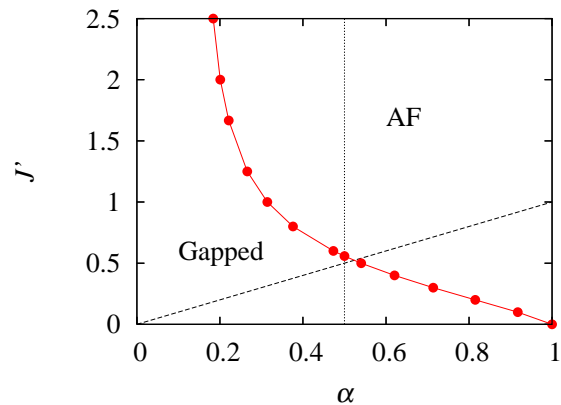}}}
  \caption{Ground state phase diagram of the non-diluted 2D $S=1/2$ AF
  Heisenberg model. The circles denote the phase boundary between the
  spin-gapped and AF LRO phases. The data points except for
  $\alpha=0.5$ were estimated by the previous 
  quantum Monte Carlo simulation~\cite{matsumoto}. The error bar of
  each point is much smaller than the symbol size. 
  The dotted and dashed lines denote
  $\alpha=0.5$ and $J'=\alpha$, respectively.}  
\label{ph-non-dil}
\end{figure}

In the present work we concentrate on the two-dimensional (2D) $S=1/2$
AF Heisenberg model consisting of coupled bond-alternating chains.
The magnitude of the stronger (weaker) intra-chain interaction is put
unity ($\alpha$) and that of the inter-chain interaction $J'$ (see also
Fig.~\ref{site-eff-J} below).  The
ground state of decoupled chains, i.e., $J'=0$, is the dimer state with
a finite spin gap, $\Delta_{\rm p}$, except at the uniform point
($\alpha=1$)~\cite{bulaevskii}, which has a critical ground state with
the Gaussian universality~\cite{affleck}.  As $J'$ increases, the
quantum fluctuations, which invoked the dimer state, are gradually
suppressed and the spin gap decreases.  When $J'$ exceeds a certain
critical value, $J'_{\rm c,p}$, which is a function of $\alpha$, the spin
gap vanishes and the AF LRO emerges.  A
global ground state phase diagram, obtained by the previous quantum
Monte Carlo simulation,~\cite{matsumoto} is presented in
Fig.~\ref{ph-non-dil}.

The ground state in the spin-gapped phase is described qualitatively
well by a direct product state of singlet dimers sitting on each strong
bond, which is called the {\em valence-bond solid state}.
When spins are randomly removed (site
dilution), a spin which formed a singlet pair formerly with a removed spin
becomes nearly free.  We call it an {\em effectively free
spin}, or simply an {\em effective spin}.  Due to the
presence of weak
interactions with the surrounding spins, an effective spin is somewhat
{\em blurred\,}; its linear extent is proportional to the correlation
length of the non-diluted system.  Between two effective spins
at sites $m$ and $n$, there exists an effective interaction, ${\tilde
J}_{mn}$, mediated by the sea of singlet pairs as illustrated in
Fig.~\ref{site-eff-J}.  It is either ferromagnetic or AF depending if
$m$ and $n$ are on the same sublattice or on the different sublattices.
It preserves its staggered nature with regard to the original square lattice
and thus the dilution does not introduce frustration.  In this sense we
call $\{{\tilde J}_{mn}\}$ the {\em effective AF interactions} between
effective spins.  Its magnitude is finite, though exponentially small,
even if $x$, the concentration of site dilution, is vanishingly small.
Consequently, we expect $x_{\rm c}^{\rm s}$, the critical concentration
of site dilution above which the disorder-induced AF LRO appears, is
strictly zero.~\cite{imada,wessel,yasuda}

\begin{figure}[t]
 \centerline{\resizebox{0.27\textwidth}{!}{\includegraphics{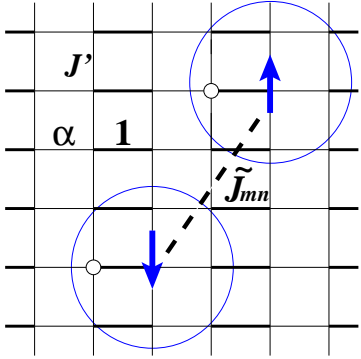}}}
 \caption{Effective spins (arrows) and an effective AF interaction
 in-between (${\tilde J}_{mn}$) in a site-diluted system. The thick and
 thin lines denote the strong and weak interactions, respectively, and
 the small circles represent the removed sites. The circles surrounding
 the arrows describe the extent of the effective spin schematically,
 which radius corresponds to the correlation length of the pure (i.e.,
 non-diluted) system.}
 \label{site-eff-J}
\end{figure}

\begin{figure}[t]
 \centerline{\resizebox{0.27\textwidth}{!}{\includegraphics{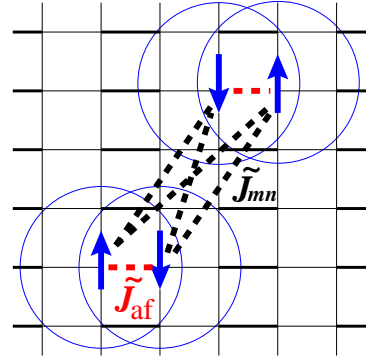}}}
 \caption{Two types of effective interactions ${\tilde J}_{mn}$ and
 ${\tilde J}_{\rm af}$ in the bond-diluted system.  The former are
 mediated by the sea of singlet pairs as the same as the site-dilution
 case, whereas the latter act through the shortest 2D paths.}
 \label{bond-eff-J}
\end{figure}

When a strong bond of amplitude unity is removed, on the other hand,
two spins on the both edges of the removed bond become nearly free as
shown in Fig.~\ref{bond-eff-J}.  We can also notice from the figure that
there exists an effective AF interaction, ${\tilde J}_{\rm af}$, between
these two spins, which is mediated mainly through the shortest paths of
the interactions connecting them, i.e., ${\tilde J}_{\rm af} \sim J'^2$.
This interaction alone recombines the two spins to form a singlet pair
with an excitation gap $\Delta$, which is proportional to ${\tilde
J}_{\rm af}$.  It yields a localized low-lying excited state, so-called
{\em mid-gap state}, as long as $\Delta \ll \Delta_{\rm p}$.
In the case of random bond dilution, in addition to ${\tilde J}_{\rm
af}$, there also exist effective AF interactions,
$\{{\tilde J}_{mn}\}$, between effective spins on edges of different
removed bonds as in the site-diluted system (Fig.~\ref{bond-eff-J}).
Naturally, we expect
competition between the two effective interactions, ${\tilde J}_{\rm
af}$ and
$\{{\tilde J}_{mn}\}$.

In the present paper, we
discuss how the competition between these two types of interactions
gives rise to the bond-dilution-induced AF
LRO associated with a finite critical concentration of bond dilution,
$x_{\rm c}\ (>0)$, in contrast to $x_{\rm c}^{\rm s}=0$
for the site dilution.  We also investigate in some detail the mid-gap
state due to ${\tilde J}_{\rm af}$ in the weak dilution limit, $x
\rightarrow 0$, and
shortly refer to a peculiar phenomenon which originates from the
randomness in the system, namely, a possible existence of the quantum
Griffiths phase neighboring to the AF LRO state.

The present paper is organized as follows. In \S2, the model and the
method of our numerical analyses are introduced. Detailed analyses of the
mid-gap state in the limit of $x \to 0$ are presented in \S3, and 
in 
\S4 the ground state phase diagram on the
$x$--$J'^2$ plane is determined. 
Section 5 is devoted to summary and
discussion.

\section{Model and Numerical Method}

We consider the bond-diluted quantum AF Heisenberg model on the square
lattice of coupled bond-alternating chains. Its Hamiltonian is
written as
\begin{eqnarray}
   \label{ham}
   {\cal H}&=&\sum_{i,j}\epsilon_{(2i,j)(2i+1,j)}
             {\bf S}_{2i,j}\cdot{\bf S}_{2i+1,j}   \nonumber \\
    &+&\alpha\sum_{i,j}\epsilon_{(2i+1,j)(2i+2,j)}
             {\bf S}_{2i+1,j}\cdot{\bf S}_{2i+2,j} \nonumber \\
    &+&J'\sum_{i,j}\epsilon_{(i,j)(i,j+1)}
             {\bf S}_{i,j}\cdot{\bf S}_{i,j+1} \ ,
\end{eqnarray}
where 1 and $\alpha$ ($>0$) are the AF intra-chain alternating coupling
constants, $J'$ ($>0$) the AF inter-chain coupling constant, and 
${\bf S}_{i,j}$ is the quantum spin operator with magnitude
$S=1/2$ at site
($i,j$). We choose the $x$-axis as the one along chains and
the $y$-axis as in the inter-chain direction.
Randomly quenched bond occupation factors \{$\epsilon_{(i,j)(k,l)}$\}
independently take either 1 or 0 with probability $1-x$ and $x$,
respectively, where $x$ is the concentration of bond dilution.

The quantum Monte Carlo (QMC) simulations with the
continuous-imaginary-time loop algorithm~\cite{MC,todo} are carried out
on $L\times L$ square lattices with periodic
boundary conditions. For each sample with a bond-diluted configuration,
$10^3 \sim 10^4$ Monte Carlo steps (MCS) are spent for measurement after
$5\times 10^2 \sim 10^3$ MCS for thermalization. For each
$x$ the random average is taken over $10^1 \sim 10^3$ samples.
As the temperature is decreased, each measured quantity, e.g.,  the
structure factor, converges to a finite value.  This reflects a finite
excitation gap due to either of an intrinsic spin gap or the finiteness
of the system.  The simulations are performed at low enough
temperatures, typically $T \simeq 10^{-3}$,
where all the physical quantities of our interest do not show any
temperature dependence and thus can be identified with those at the
ground state.

\begin{figure}[t]
 \centerline{\resizebox{0.46\textwidth}{!}{\includegraphics{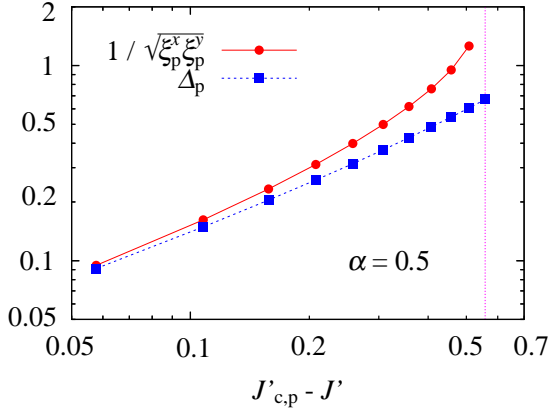}}} 
 \caption{$J'$ dependences of the the correlation lengths and the spin gap
 $\Delta_{\rm p}$ in the non-diluted system with
 $\alpha=0.5$, where $J'_{\rm c,p}=0.5577(1)$ is the critical value of
 $J'$ determined by the same method as in ref.~\citen{matsumoto}.
 The dotted line denotes $J'=0$, the system of decoupled chains.
 The error bar of each point is
 smaller than the symbol size. The solid and dashed lines going through the
 data points are guides
 to eyes.} 
 \label{corr-gap-pure}
\end{figure}

Before going into discussions on bond-diluted systems, let us here
mention some ground state properties of the non-diluted 
systems described by Hamiltonian (\ref{ham}) with $x=0$.
Its ground state phase diagram on the $\alpha$--$J'$ plane is shown in
Fig.~\ref{ph-non-dil}. For discussions below we show in 
Fig.~\ref{corr-gap-pure} the $J'$ dependence of the spin gap 
$\Delta_{\rm p}$ and the inverse geometric mean of the
correlation lengths $\xi_{\rm p}^{x}$ and $\xi_{\rm p}^{y}$ along
$\alpha=0.5$, on which we concentrate our analyses of
bond-dilution effects below. 
The correlation lengths $\xi_{\rm p}^{x}$ and $\xi_{\rm p}^{y}$ are
estimated from the dynamical correlation functions at momenta $(\pi,0)$
and $(0,\pi)$, respectively, by using the second-moment
method~\cite{cooper}. Similarly, $\Delta_{\rm p}$ is evaluated from the
correlation function along the imaginary-time axis (see also below).  
As seen in Fig.~\ref{corr-gap-pure}, both $\Delta_{\rm p}^{-1}$ 
and $\xi_{\rm p}^{x} \xi_{\rm p}^{y}$
exhibit power law divergence as $J'$ approaches $J'_{\rm c,p}$ from below.

\section{Mid-Gap State in the Limit of $x \to 0$}

When a strong bond is removed from the non-diluted system with the
non-magnetic ground state, two magnetic moments
induced at the both ends of the diluted bond are expected to reform a spin
singlet through ${\tilde J}_{\rm af}$, one of the effective AF
interactions mentioned in \S1.  For a sufficiently weak bond dilution,
$x \ll 1$, we expect a mid-gap state to appear, since the spin gap
$\Delta$ induced by ${\tilde J}_{\rm af}$ is smaller than $\Delta_{\rm
p}$, the spin gap in the non-diluted system.  In this section, based on
the QMC data,
we discuss the $J'$ dependence of the gap attributed to
${\tilde J}_{\rm af}$ in the limit of $x \to 0$.

\begin{figure}[t]
 \centerline{\resizebox{0.46\textwidth}{!}{\includegraphics{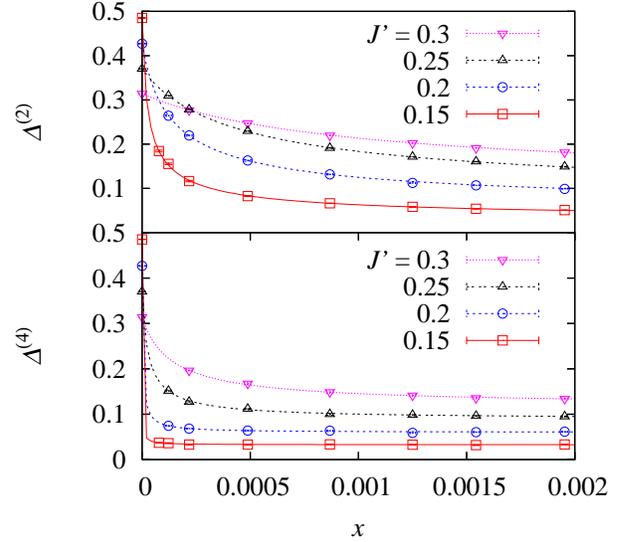}}}
 \caption{Dependences of $\Delta^{(2)}$ and $\Delta^{(4)}$ estimated by
 the second- and fourth-moment estimators on $x=1/(2L^2)$ for
 $\alpha=0.5$ and $L\times L$ systems with 
 $L=16$, 18, 20, 24, 32, 48, 64, and 80. The lines are the fitting
 functions [eqs.~(\ref{fit2}) and (\ref{fit4})].}
 \label{midgap-fit}
\end{figure}

To extract the value of $\Delta$, we make use of the second-moment
formula,~\cite{todo,cooper}
\begin{equation} 
 \label{xi2}
 (\Delta^{(2)})^{-1}=\frac{\beta}{2\pi}\sqrt{\frac{\tilde{C}(0)}{\tilde{C}(2\pi/\beta)}-1} \ .
\end{equation}
Here $\beta$ is the inverse of temperature and
$\tilde{C}(\omega)=\int_{0}^{\beta}{\rm d}\tau C(\tau)e^{i\omega\tau}$
is the Fourier transform of the imaginary-time staggered correlation
function
\begin{equation}
   C(\tau) \equiv \frac{1}{L^{2}}\sum_{i,j}
         (-1)^{|r_{i}-r_{j}|}
        \langle S_{i}^{z}(0)S_{j}^{z}(\tau) \rangle \ .
\end{equation}
In the bond-diluted system of present interest, it is expected that
$C(\tau)$ is dominated by two terms; the one representing the original
spin gap in the non-diluted system and the one attributed to the mid
gap.  That is, the imaginary-time staggered correlation function can be
approximated as~\cite{yasuda3,errata}
\begin{eqnarray}
 \label{cosh}
  C(\tau) &=& a_{\rm p} \Big[ {\rm cosh}\frac{\beta\Delta_{\rm p}}{2} \Big]^{-1}
            {\rm cosh}\{(\tau-\beta/2)\Delta_{\rm p}\} \nonumber \\
          &+& a_{\rm i} \Big[ {\rm cosh}\frac{\beta\Delta}{2} \Big]^{-1}
            {\rm cosh}\{(\tau-\beta/2)\Delta\} \ ,
\end{eqnarray}
where $a_{\rm p}$ and $a_{\rm i}$ are some functions of $\alpha$, $J'$,
and $x$. The coefficient $a_{\rm i}$, which represents the {\em weight} of the
contribution from the mid-gap state, is naturally considered to be
proportional to the concentration of
diluted bonds, i.e., $a_{\rm i}= a x$ with $a = a(\alpha, J')$.  From
eqs.~(\ref{xi2}), (\ref{cosh}), and $a_{\rm p}+a_{\rm i}=1$,
$\Delta^{(2)}$ is evaluated as
\begin{equation}
 \label{fit2}
\Delta^{(2)}=\Delta\sqrt{(1+\frac{1-ax}{ax}\frac{\Delta}{\Delta_{\rm
p}})/\{{1+\frac{1-ax}{ax}(\frac{\Delta}{\Delta_{\rm p}})^3}\}}
\end{equation}
for $\beta^{-1} \ll \Delta \ll \Delta_{\rm p}$.  Note that in the last expression,
$\Delta^{(2)}$ explicitly depends on $x$ as well as on $\Delta$.  We
regard eq.~(\ref{fit2}) as a fitting function to extract $\Delta$ in the
weak dilution limit, where $\Delta$ is assumed to be independent of $x$.

\begin{figure}[t]
 \centerline{\resizebox{0.46\textwidth}{!}{\includegraphics{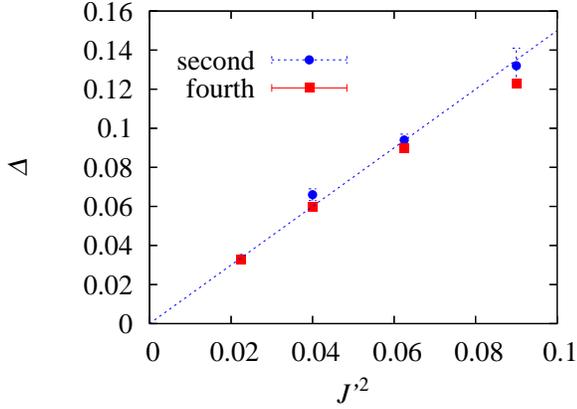}}}
 \caption{$J'^2$ dependences of the spin gap estimated by the second-
 and fourth-moment estimators. The dashed line, which represents $\Delta
 = 1.5J'^2$, is a guide to eyes.}
 \label{midgap}
\end{figure}

The QMC simulation to evaluate $\Delta^{(2)}$ is performed along
$\alpha=0.5$ for the system with only one diluted bond, which
corresponds to the $x=1/2L^2$ system. By taking the limit of $L \to
\infty$, we can investigate the properties in the limit of $x \to 0$.
The obtained set of $\Delta^{(2)}(x)$ for each $J'$ is fitted by
eq.~(\ref{fit2}), thereby the value of
$\Delta_{\rm p}$ is fixed to the one simulated in the non-diluted
system.  In the upper part of Fig.~\ref{midgap-fit} the $x$ dependence
of $\Delta^{(2)}$ is presented.
The resultant spin gap $\Delta$ is shown in Fig.~\ref{midgap}. As a
function of $J'$, it increases proportionally to $J'^2$, while
$\Delta_{\rm p}$ decreases (see Fig.~\ref{corr-gap-pure}).

We have further confirmed the values of $\Delta$
by using the fourth-moment estimator~\cite{todo}, defined by
\begin{equation}
 \label{xi4}
(\Delta^{(4)})^{-1}=\frac{\beta}{4\pi}\sqrt{3\frac{\tilde{C}(0)-\tilde{C}(2\pi/\beta)}{\tilde{C}(2\pi/\beta)-\tilde{C}(4\pi/\beta)}-1} \ .
\end{equation}
In this case the fitting function becomes
\begin{equation}
 \label{fit4}
 \Delta^{(4)}=\Delta\sqrt{\{{1+\frac{1-ax}{ax}(\frac{\Delta}{\Delta_{\rm p}})^3\}/\{1+\frac{1-ax}{ax}(\frac{\Delta}{\Delta_{\rm p}})^5}\}} \ .
\end{equation}
The lower part of Fig.~\ref{midgap-fit} is the $x$
dependence of $\Delta^{(4)}$ obtained by eq.~(\ref{xi4}).
It is observed that as $x$ increases $\Delta^{(4)}$ converges more
rapidly to a constant values, which corresponds to $\Delta$.  This is
due to the smaller corrections, $[(1-ax)/ax](\Delta/\Delta_{\rm
p})^3$, in eq.~(\ref{fit4}) than those for $\Delta^{(2)}$.

Figure~\ref{midgap} shows a nice coincidence of the values
of $\Delta$, estimated by the second- and fourth-moment
methods. 
We therefore conclude the existence of a mid-gap state
whose gap $\Delta$ is proportional to $J'^2$ and distinctly smaller than
$\Delta_{\rm p}$, and whose contribution to the
imaginary-time correlation function is proportional to $x$.
This is just what we have expected, i.e., a reformed singlet
pair due to the effective AF interaction ${\tilde J}_{\rm af}$ acting on
two effective spins at both ends of a diluted strong bond.

\section{Bond-Dilution-Induced Antiferromagnetic Long-Range-Ordered Phase}

\begin{figure}[t]
 \centerline{\resizebox{0.46\textwidth}{!}{\includegraphics{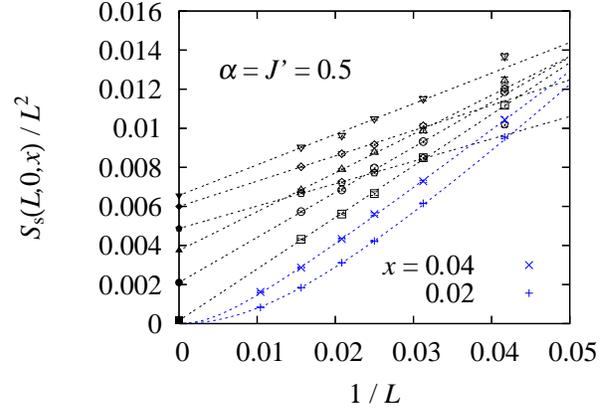}}}
 \caption{System size dependences of $S_{\rm s}(L,0,x)/L^2$ for $x=0.02$
 (crosses), 0.04 (x-marks), 0.06 (squares), 0.08 (circles), 0.1 (upward
 triangles), 0.2 (downward triangles), 0.3 (diamonds), and 0.35
 (pentagons) at $(\alpha,J')=(0.5,0.5)$.  Dashed lines are obtained by
 the least-squares fitting. The extrapolated values are denoted by
 the filled symbols which correspond to $M_{\rm s}^2/3$ in the thermodynamic
 limit.}  \label{sstrL}
\end{figure}

\begin{figure}[t]
 \centerline{\resizebox{0.46\textwidth}{!}{\includegraphics{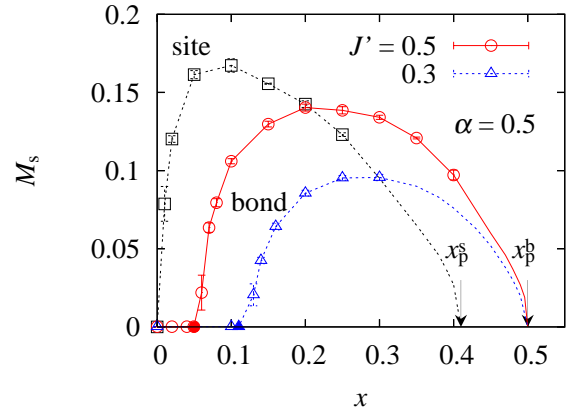}}}
 \caption{Dependences of the staggered magnetization on the concentration
 of bond dilution for $(\alpha,J')=(0.5,0.5)$ and $(0.5,0.3)$.
 The solid circle and triangle denote the quantum critical
 points, $x_{\rm c} \approx 0.0502$ and 0.1101, respectively, estimated by the
 finite-size scaling analysis described in the text.  
For comparison, we also show the result of the site dilution
 for $(\alpha,J')=(0.5,0.5)$, which is taken from ref.~15. 
The percolation thresholds
 on the site and bond processes are denoted by 
$x_{\rm p}^{\rm s} \approx 0.41$ and $x_{\rm p}^{\rm b}=0.5$,
 respectively. All the lines are guides to eyes.}
\label{ms_fig}
\end{figure}

\begin{figure*}[t]
 \centerline{%
 \begin{minipage}[b]{.2cm}(a)\vspace*{4.9cm}\end{minipage}
 \hspace*{-.5cm}
 \resizebox{0.46\textwidth}{!}{\includegraphics{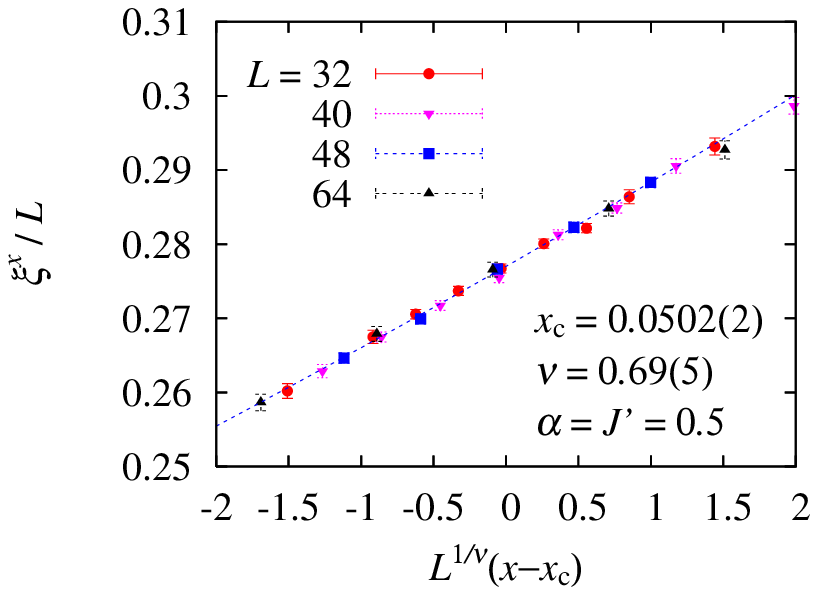}}
 \hspace*{.5cm}
 \begin{minipage}[b]{.2cm}(b)\vspace*{4.9cm}\end{minipage}
 \hspace*{-.5cm}
 \resizebox{0.46\textwidth}{!}{\includegraphics{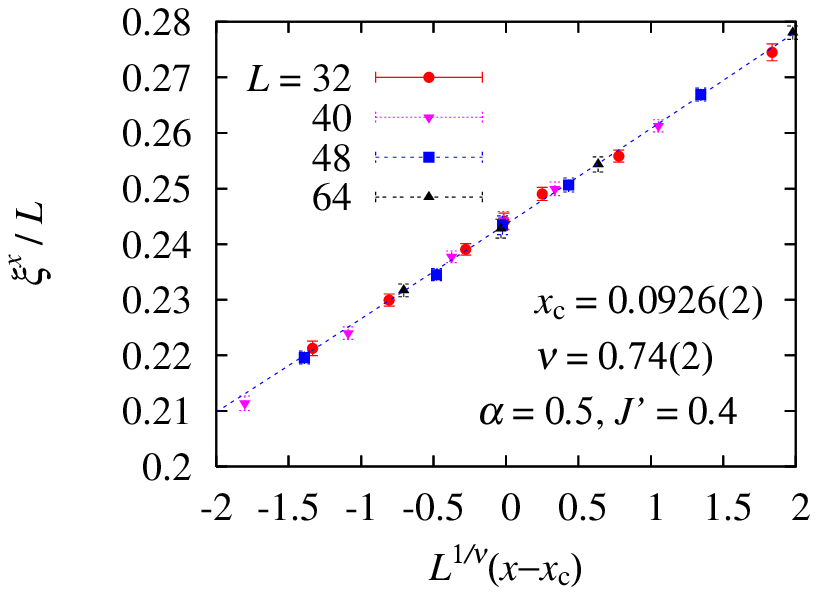}}
 }

 \centerline{%
 \begin{minipage}[b]{.2cm}(c)\vspace*{4.9cm}\end{minipage}
 \hspace*{-.5cm}
 \resizebox{0.46\textwidth}{!}{\includegraphics{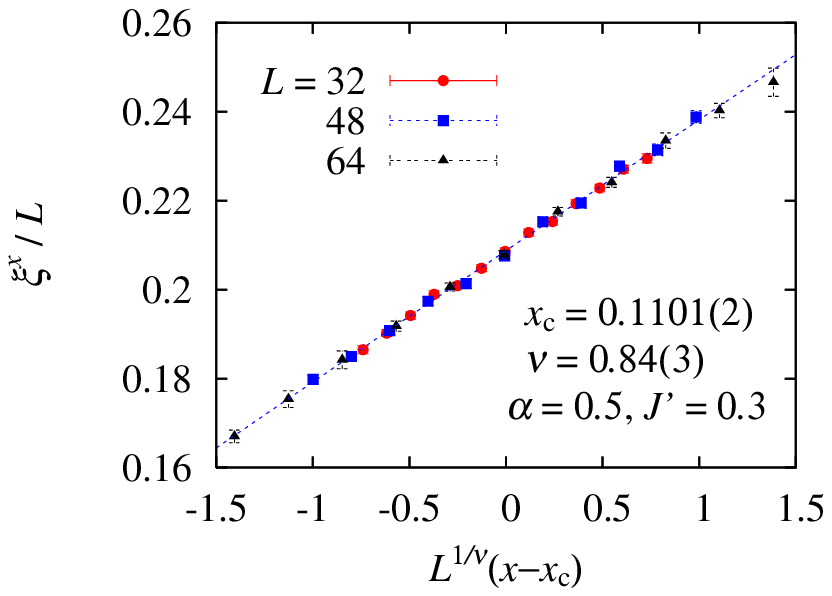}}
 \hspace*{.5cm}
 \begin{minipage}[b]{.2cm}(d)\vspace*{4.9cm}\end{minipage}
 \hspace*{-.5cm}
 \resizebox{0.46\textwidth}{!}{\includegraphics{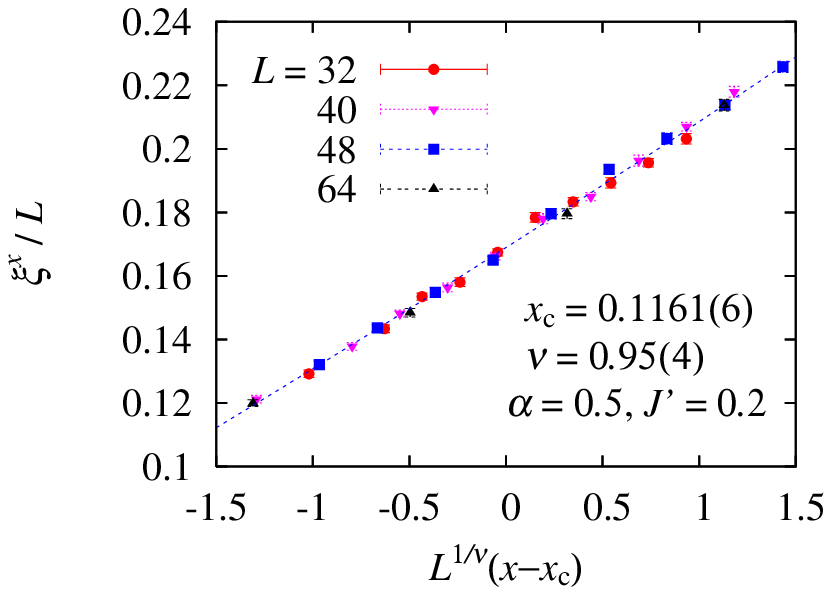}}
 }

 \caption{Finite-size scaling plots of the correlation length for
 (a) $(\alpha,J')=(0.5,0.5)$, (b) $(0.5,0.4)$, (c) $(0.5,0.3)$, and (d)
 $(0.5,0.2)$. }
 \label{corr-scale}
\end{figure*}

As $x$ is increased from zero, fitting of $C(\tau)$ by using
eq.~(\ref{cosh}) becomes unstable, implying that
a number of mid-gap states, other than $\Delta$ discussed in the previous
section, appear. 
We have clearly observed, however, that $\Delta^{(2)}$ of
eq.~(\ref{xi2}), which we now regard as an upper bound of the spin gap,
as well as the inverse of the spatial correlation lengths, $\xi^x$ 
and $\xi^y$, decrease
as $x$ increases.~\cite{yasuda3} Finally, when $x$ exceeds a certain
critical value, $x_{\rm c}$, the AF LRO appears.
During the present section, we call the state between $x=0$ and $x_{\rm
c}$ simply the disordered 
state, and concentrate on the quantum phase transition at $x=x_{\rm c}$ to 
the AF LRO state, postponing the discussion about a possible
existence of the quantum Griffiths state 
at $x < x_{\rm c}$ until the last section. 

To discuss the quantum phase transition between the disordered and the
AF LRO phases, first we investigate the zero-temperature staggered
magnetization, $M_{\rm s}$, as a function of $x$, which is evaluated as
\begin{equation}
   M_{\rm s}^{2}(x)=\lim_{L\to \infty}\lim_{T\to 0}
       \frac{3S_{\rm s}(L,T,x)}{L^{2}} \ .
\label{ms}
\end{equation}
Here $S_{\rm s}(L,T,x)$, the static structure factor at momentum
$(\pi,\pi)$, is defined by
\begin{equation}
    \label{str}
    S_{\rm s}(L,T,x) \equiv \frac{1}{L^{2}}\sum_{i,j}
        (-1)^{|r_{i}-r_{j}|}
        \langle S_{i}^{z}S_{j}^{z} \rangle \ .
\end{equation}
In Fig.~\ref{sstrL} we show the system size dependence of $S_{\rm
s}(L,0,x)/L^2$ for various $x$'s.  For $x \ge 0.06$, the QMC data for
each $x$ are fitted fairly well by a linear expression, $a + bL^{-1}$,
where the value of $a$, represented by the solid symbols in the figure,
gives an estimate for $M_{\rm s}^{2}/3$ in the thermodynamic limit.  For
$x=0.02$ and 0.04, on the other hand, we find a linear fit extrapolates
to a negative $M_{\rm s}^2/3$ for $L \to \infty$, which indicates a
vanishing staggered magnetization.  Indeed, the data are fitted much
better by $S_{\rm s}(L,0,x) = a (1-{\rm exp}(-b L))$, which
is the scaling form derived for a non-magnetic ground state in the
modified spin-wave theory.~\cite{modified_spin_wave}

In Fig.~\ref{ms_fig} the $x$ dependences of $M_{\rm s}$ are shown for
$J'=0.3$ and 0.5, together with that for the site-diluted
system~\cite{yasuda} for comparison. The value of $\alpha$ is fixed to
be 0.5 in all the cases.
It is clearly demonstrated that
while the AF LRO is induced at an infinitesimal concentration of site
dilution, in the bond-diluted system there exists 
a critical concentration, e.g., $x_{\rm c} \approx 0.11$ for $J'=0.3$. 
It is also definitely finite even for $J'=0.5$, whose
non-diluted system is close to the critical point 
$J'_{\rm c,p} \approx 0.5577$.

In passing we note that 
$M_{\rm s}(x)$ shown in Fig.~\ref{ms_fig} has a peak at
$x=x_{\rm opt}$. In the strong dilution regime, 
$x_{\rm opt} \lesssim x \le x_{\rm p}^{\rm s,b}$, 
the further dilution tends to destroy the AF LRO and 
the latter would vanish
just at the percolation threshold~\cite{Kato, TodoYKHKMT2000}.  

Next let us determine the phase diagram on the $x$--$J'^2$ plane.
The critical concentration of dilution, $x_{\rm c}$, which is roughly
evaluated above, is estimated more precisely by the
finite-size scaling analysis of the correlation length 
$\xi^x$ as shown in Fig.~\ref{corr-scale}.
The critical concentrations are estimated as $x_{\rm c}= 0.0502(2)$,
0.0926(2), 0.1101(2), and 0.1161(6) for $J'=0.5$, 0.4, 0.3, and 0.2,
respectively.  [Almost the equal values of $x_{\rm c}$ are
obtained by the same analysis on $\xi^y$.]
The values of the critical exponent of the correlation
length are $\nu=0.69(5)$, 0.74(2), 0.84(3), and 0.95(4), respectively,
which monotonically increases from $\nu=0.71$, the critical exponent of
the three-dimensional
classical Heisenberg universality class.~\cite{Chen}
We note, however, that the finite-size scaling analysis with the
fixed critical exponent $\nu=0.71$ gives the same value of
$x_{\rm c}$ within the numerical accuracy. Certainly, more precise
numerical data are needed to settle this point.

\begin{figure}[t]
 \centerline{\resizebox{0.46\textwidth}{!}{\includegraphics{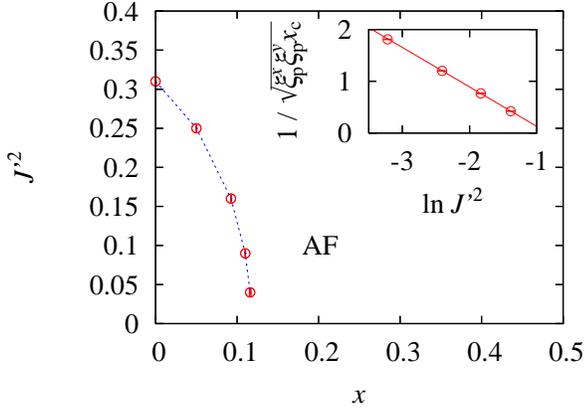}}}
 \caption{Phase diagram of the ground state for $\alpha=0.5$. The
 circles are evaluated by the finite-size scaling analysis. The dashed 
 line is a guide to eyes. In the inset we show the relation between
 $1/\sqrt{\xi_{\rm p}^x \xi_{\rm p}^y x_{\rm c}}$ and $\ln{J'^2}$, where
 we used the QMC data ($\xi_{\rm p}^x, \xi_{\rm p}^y$) = (11.998(9),
 9.312(10)), (5.320(1), 3.460(2)), (3.452(1), 1.8202(3)), and (2.5492(4),
 1.03244(5)) for $J'=0.5$, 0.4, 0.3, and 0.2, respectively. The straight
 line in the inset represents
 a linear fitting of the data by $y=a+bx$ with $a=-0.633(6)$ and $b=-0.762(3)$.}
 \label{j2-x-2}
\end{figure}

The resultant phase diagram is shown in Fig.~\ref{j2-x-2}. 
In the inset of the figure we show an important result, i.e., the
relation of $x_{\rm c}$ to $J'^2$, or the inverse of
the function $J'_{\rm c}(x)$, a critical value of $J'$ for a given $x$
(and a given $\alpha$). Note that $J'_{\rm c,p} = J'_{\rm c}(x=0)$.
The effective AF interaction, $\tilde{J}_{mn}$, introduced in 
\S1 is approximately described by 
\begin{equation}
  \tilde{J}_{mn} \propto (-1)^{|r_m-r_n+1|} {\rm exp}(-\frac{\ell}{\sqrt{\xi_{\rm p}^x \xi_{\rm p}^y}}) \ , 
 \label{jmn}
\end{equation}
where $\ell=|r_m-r_n|$ is the distance between the effective 
spins~\cite{nagaosa,iino} at the ends of {\it different} removed bonds. 
On the other hand, the effective AF interaction, $\tilde{J}_{\rm af}$,
between effective spins on the same removed bond is simply described by
\begin{equation}
 \tilde{J}_{\rm af} \simeq J'^2 \ .
 \label{jaf}
\end{equation}
Thus the relation indicated in the inset of Fig.~\ref{j2-x-2} is read as
\begin{equation}
  {\cal J}(x_{\rm c}) \simeq \tilde{J}_{\rm af} \ ,
\label{calJ}
\end{equation}
where
${\cal J}(x)$ is an average of $\{{\tilde J}_{mn}\}$ acting
on effective spins on neighboring diluted bonds, and is
approximately represented as
\begin{equation}
  {\cal J}(x) \simeq {\cal J}_0 \ {\rm exp}(-\frac{1}{\sqrt{\xi_{\rm p}^x \xi_{\rm p}^y x}}) \ ,
\label{calJ-2}
\end{equation}
in which the exponential part is obtained from $\tilde{J}_{mn}$ of
eq.~(\ref{jmn}) with $1/\sqrt{x}$, i.e., the mean distance of
the nearest removed bonds, substituted for $\ell$. 
The result of eq.~(\ref{calJ}) implies
that the phase transition between the 
disordered and AF LRO phases occurs when the magnitudes of $\tilde{J}_{\rm af}$
and ${\cal J}(x)$ become comparable.
We interpret this result as described below.

Let us replace a bond-diluted system of present interest by a model system
consisting only of effective spins 
with interactions $\{{\tilde J}_{mn}\}$ and ${\tilde J}_{\rm af}$,
neglecting all spins in the sea of singlet pairs. Since 
$\{{\tilde J}_{mn}\}$ do not introduce frustration at all, 
combined with the fact that $\{{\tilde J}_{mn}\}$ are short-ranged
except for the system with $J'=J'_{\rm c,p}$, the actual geometry of  
$\{{\tilde J}_{mn}\}$ is not relevant to the present problem, and so we
may further simplify the model system to a regular square array of the
effective spins; along the 
chain direction the interactions are
alternating between ${\tilde J}_{\rm af}$ and 
${\cal J}(x)$, while
in the inter-chain direction the interaction is uniform, i.e., also 
${\cal J}(x)$ given above. 
In this way, the original system in the disordered state can be
approximately mapped to a gapped state of the original non-diluted system
of Fig.~\ref{ph-non-dil} but with both $\alpha$ and $J'$ replaced by 
${\cal J}(x) / \tilde{J}_{\rm af}$. As $x$ (and/or $J'$) of the
bond-diluted system increases, 
${\cal J}(x) / {\tilde J}_{\rm af}$ does so,
which corresponds to the upward movement of the mapped regular system
along the dashed line ($J' = 
\alpha$) in the figure. 
Finally it hits the critical line at $x=x_{\rm c}(J')$ 
($J'=J'_{\rm c}(x)$),
where eq.~({\ref{calJ}}) holds, and the quantum phase transition to the 
AF LRO phase occurs. 

Our above scenario can explain the behavior of the
critical line in the $x \rightarrow 0$ limit, i.e., 
$J'_{\rm c}(x)$ at least first decreases from 
$J'_{\rm c,p}$ when the dilution is introduced  
(see Fig.~\ref{j2-x-2}).  
Naively one might expect the dilution suppresses the bulk correlation
represented by $\xi_{\rm p}^{x,y}$, and in order to recover this decrease, 
 the stronger $J'$ is needed for
the AF LRO to appear, yielding an initial increase of 
$J'_{\rm c}(x)$.
If, however, our scenario is applied to the system just at
$J'=J'_{\rm c,p}$, where the correlation lengths $\xi_{\rm p}^{x,y}$ are
infinite, it is expected that the AF LRO is induced by an infinitesimal
concentration of dilution, since $\{J_{mn}\}$ is quasi-long-ranged even
for a vanishingly small $x$ and the prefactor in
eq.~(\ref{calJ-2}), ${\cal J}_0$, depends more weakly on $J'$ than
${\tilde J}_{\rm af}\ (\propto J'^2)$ does.
This means that $J'_{\rm c}(x)$ has a negative slope at
least at $x=0^+$. Actually, from eqs.~(\ref{calJ}) and (\ref{calJ-2})
combined with the critical behavior of $\xi_{\rm p}^{x,y}$ at
$J'\simeq J'_{\rm c,p}$, the decrease of $J'_{\rm c}(x)$ at 
$x \sim 0$ is deduced. However, quantitative details of its behavior in the
weak dilution limit are beyond a scope of the present work.

\section{Conclusion and Discussion}

In the present paper, we have investigated the bond-dilution effects on the 
non-magnetic ground state of the 2D quantum AF Heisenberg model
consisting of bond-alternating chains by means of the QMC
simulations, and have proposed a scenario that the
quantum phase transition induced
by bond dilution is originated from the competition of the two
effective AF interactions $\{{\tilde J}_{mn}\}$ and 
${\tilde J}_{\rm af}$. 
In particular, the proposed simple mapping of a
bond-diluted system to a non-diluted system catches up the essential
mechanism of the bond-dilution-induced AF LRO in the system 
so long as the effective interaction ${\cal J}(x)$, which represents an
averaged value of $\{{\tilde J}_{mn}\}$, is less than or
comparable to ${\tilde J}_{\rm af}$, or in other words, $J'$ is not too
small.

In our arguments so far described, we have kept away from an interesting
problem whether the disordered state in the present system may involve
a novel aspect of quantum random systems such as the quantum Griffiths
(QG) phase or not.
By our scenario with the averaged interaction ${\cal J}(x)$ we completely
neglect this possibility.
For $J'=0$, i.e., the system is exactly one-dimensional, 
the system with $0<x<1$ is known to be
in the QG phase characterized by finite
correlation lengths but with a continuous distribution of spin gaps up
to zero.~\cite{fisher,hyman,hida,todo3}. 
If this QG state is unstable against the AF LRO state due to the
introduction of $J'$ even of an infinitesimal magnitude, the latter
state invades to $x \rightarrow 0$, i.e., a reentrant transition from 
the disordered state to the AF LRO state as decreasing $J'$ is expected 
for $x$ smaller than 0.11. Also for a moderate value of $J'$ for which 
we have observed the
quantum phase transition from the disordered state to the AF LRO state, 
there remains the problem of a possible existence of the QG state. 
Actually in our preliminary analysis we observed 
in the $J'=0.3$ system that $\Delta^{(2)}$ of 
eq.~(\ref{xi2}) decreases much faster than the inverse of
the spatial correlation length as $x$ increases.~\cite{yasuda3} 
This strongly suggests an existence of the QG state. 
However, in order to confirm a continuous distribution of spin gaps up
to zero, we have to carry out QMC simulations at correspondingly low 
temperatures.
This problem as well as that of the possible reentrant transition
mentioned above remain as challenging future works in the field
of the computational physics.

In order to distinguish experimentally the characteristic difference
between the site- and bond-dilution-induced AF LRO states pointed out in
the present work, one needs a suitable material; a quasi-one-dimensional
Heisenberg antiferromagnet with small enough $J'$ and rigid bond
alternation, and without frustration. Furthermore these properties are
desired not to change significantly by dilution processes.  
In the spin-Peierls compound CuGe$_{1-x}$Si$_{x}$O$_3$, which is
effectively a bond-disorder Heisenberg
antiferromagnet~\cite{regnault}, it was experimentally observed that the
effective spins are induced not near Si impurity sites
but rather at sites between them in the AF LRO
phase~\cite{kikuchi}. The occurrence of the AF LRO is then attributed to
the nonlocal reconstruction of bond alternation by Si substitution. To
understand this type of phase transitions in the spin-Peierls state, we 
have to analyze models with proper spin-lattice coupling, which will be
another challenging numerical work in near future.

\section*{Acknowledgment}

The authors acknowledge M. Matsumoto for stimulating discussions.
Most of the numerical calculations were performed on the SGI 2800 at
Institute for Solid State Physics, University of Tokyo. The program is
based on `Looper version 2' developed by S.T. and K. Kato~\cite{looper}
and `PARAPACK
version 2' by S.T.  This work is partially supported by Grants-in-Aid
for Scientific Research Programs (Nos.~15740232, 18540369, and 18740239)
and that for the 21st
Century COE Program from the Ministry of Education, Culture, Sports,
Science and Technology of Japan.


\begin{thebibliography}{99}
\bibitem{errata} The prefactors in eq.~(3.4) in ref.~\citen{yasuda3}
	should be read respectively for $(\cosh \beta \Delta_{\rm p} /
	2)^{-1}$ and $(\cosh \beta \Delta / 2)^{-1}$.  The following
	equations and discussions in ref.~\citen{yasuda3} are not
	affected by these typographic errors.
\end{thebibliography}
\end{document}